\documentclass[doublespace,fleqn,10pt]{wlscirep}
\title{High-throughput computation and structure prototype analysis for two-dimensional ferromagnetic materials}
\author[1,2]{Zhen-Xiong Shen}
\author[1,2,*]{Chuanxun Su}
\author[1,2,**]{Lixin He}
\affil[1]{Key Laboratory of Quantum Information, University of Science and
Technology of China, Hefei, 230026, China}
\affil[2]{Synergetic Innovation Center of Quantum Information and Quantum
  Physics, University of Science and Technology of China, Hefei, 230026, China}

\affil[*]{sucx@ustc.edu.cn}
\affil[**]{helx@ustc.edu.cn}

\usepackage{setspace}%
\usepackage{graphicx}
\usepackage{float}
\usepackage{dcolumn}
\usepackage{color}
\usepackage{latexsym,bm}
\usepackage[normalem]{ulem}
\usepackage{multirow}
\usepackage{appendix}
\usepackage{epstopdf}
\usepackage{cmap}
\usepackage{gensymb}
\usepackage{url}
\usepackage{soul}

\begin{abstract}
We perform high-throughput first-principles computations to search the high Curie temperature ($T_{\rm C}$)  two-dimensional ferromagnetic (2DFM) materials. We identify 79 2DFM materials and calculate their $T_{\rm C}$, in which  Co$_2$F$_2$ has the highest $T_{\rm C}$=541K, well above the room temperature.
The 79 2DFM materials are classified into different structural prototypes according to their structural similarity.
We perform  sure independence screening and sparsifying operator (SISSO) analysis to explore the relation
between $T_{\rm C}$ and the material structures. The results suggest that the 2DFM materials with shorter distance between the magnetic atoms, larger local magnetic
moments and more neighboring magnetic atoms are more likely to have higher $T_{\rm C}$.
\end{abstract}

\begin{document}
\begin{spacing}{2.0}

\flushbottom
\maketitle

\thispagestyle{empty}

\section*{Introduction}
Long-range magnetic order is suppressed in two-dimensional (2D) isotropic systems at finite temperatures due to
thermal fluctuations, according to the Mermin-Wagner theorem~\cite{Mermin1966Absence}.
Therefore, recent discoveries of ferromagnetism in 2D materials, e.g.,
CrI$_{3}$~\cite{Huang2017Layer,Thiel2019Probing}, Cr$_{2}$Ge$_{2}$Te$_{6}$~\cite{Gong2017Discovery}, Fe$_{3}$GeTe$_{2}$~\cite{Fei2018Two,Deng2018Gate}, and Fe$_{4}$GeTe$_{2}$~\cite{Seo2020Nearly} have attracted broad attention.
The 2D ferromagnetic (2DFM) materials  possess some fascinating in many aspects~\cite{Deng2018Gate,Gong2019Two,Sun2019Giant},
which also have great potential in device applications, such as 2D spintronics~\cite{wang2018very,song2018giant,klein2018probing,kim2018one,wang2018tunneling}.

However, only very few  2DFM materials have been experimentally synthesized so far, and the Curie temperatures ($T_{\rm C}$s)
of these 2DFM materials are fairly low.
The searching for 2DFM materials with high  $T_{\rm C}$ via first-principles calculations attracts more and more attention.
 Mounet {\it et al} applied density functional theory (DFT) calculations to search for easily exfoliable magnetic compounds~\cite{Marzari2018Two},
and they revealed a wealth of 2D magnetic systems including 37 ferromagnets. Zhu {\it el at} found 15 2DFM materials~\cite{Zhu2018Systematic}.
Especially, Cr$_{3}$Te$_{4}$  was predicted to have an extremely high $T_{\rm C}$=2057 K.
Torelli {\it et al} discovered 85 2DFM materials by high-throughput computation~\cite{Daniele2020High}.
Kabiraj et al. found 26 2DFM materials with $T_{\rm C}$ higher than 400 K from high-throughput scanning of 786 materials~\cite{Kabiraj2020High}.

In this work, we carry out a comprehensive high-throughput computational study to calculate the $T_{\rm C}$s of 2DFM materials.
The magnetic exchange interactions are calculated via a first-principles linear response theory~\cite{Liechtenstein1987Local,Bruno2003Exchange,Wan2006Calculation,Wan2009Calculated}.
We simulate the magnetic phase diagrams and evaluate the $T_{\rm C}$ by a replica-exchange Monte Carlo simulation~\cite{Cao2009First}. We have identified 79 2DFM materials.
We benchmark the calculated $T_{\rm C}$s with those of experimental synthesized materials, and also with the corresponding bulk materials.
The results suggest that the results are highly reliable.
The  compound with the highest  $T_{\rm C}$ we predicted is Co$_2$F$_2$, which has  $T_{\rm C}$=541K.
The 79 2DFM materials are classified into different structural prototypes according to their structural similarity.
We perform  sure independence screening and sparsifying operator (SISSO) analysis~\cite{fan2008sure,ouyang2018sisso} to explore the relation
between $T_{\rm C}$ and the material structures. The results suggest that the 2DFM materials with shorter distance between the magnetic atoms, larger local magnetic
moments and more neighboring magnetic atoms are more likely to have higher $T_{\rm C}$.

\section*{Results}
\subsection*{Two-dimensional magnetic atomic structural prototype}

We select 198 2D magnetic materials which have high dynamical and thermodynamic stability in the Computational 2D Materials Database (C2DB)~\cite{Sten2018The,C2DB-web,Gjerding2021Recent}.
We also calculate 37 compounds from Ref.~\cite{Marzari2018Two}, and 8 compounds from Ref.~\cite{Zhu2018Systematic}. In addition to these materials, we also include several experimentally synthesized structures such as CrI$_{3}$~\cite{Huang2017Layer,Thiel2019Probing}, Cr$_{2}$Ge$_{2}$Te$_{6}$~\cite{Gong2017Discovery}, Fe$_{3}$GeTe$_{2}$~\cite{Fei2018Two,Deng2018Gate}, and Fe$_{4}$GeTe$_{2}$~\cite{Seo2020Nearly}.
The workflow and calculation details are presented in Sec.\ref{sec:method}.
There are some 2D materials that show complicated magnetic ground states, which will be interesting for further studies. In this work, we focus on the FM materials,
which have  great potential in device applications.

\begin{figure}
  \centering
  \includegraphics[width=0.45\textwidth]{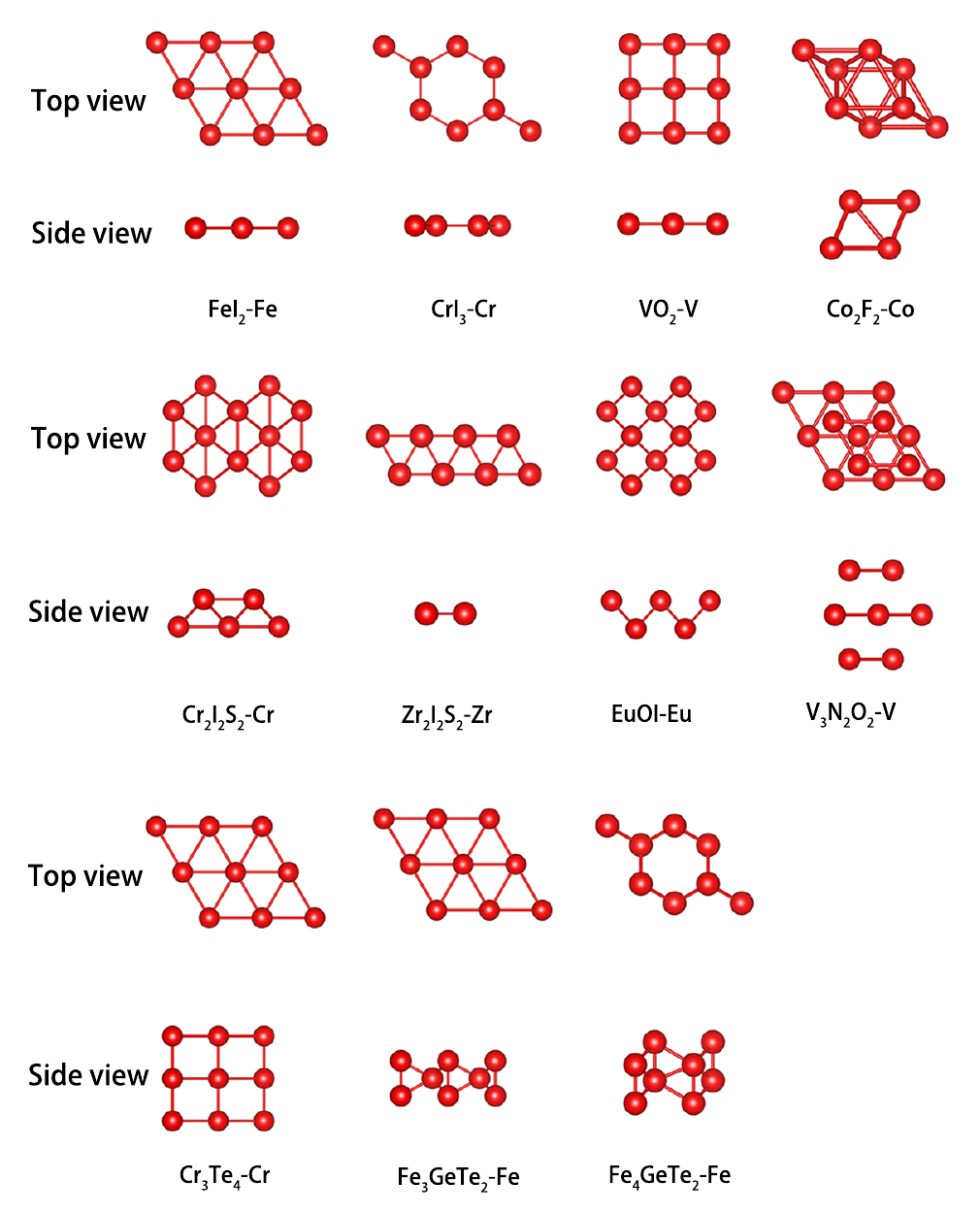}
  \caption{The structure prototypes of 2DFM materials with only magnetic atoms kept. }
  \label{fig:StruturePrototype}
\end{figure}

We find 79 2D magnetic materials that have robust FM ground states.
We analyze the structural characters of  these 2DFM materials. For simplicity, we first consider only the magnetic atoms in the unit cell.
The 79 2DFM structures can be divided into 11 categories according to the structural similarity of the magnetic atoms.
The prototypical structures of the  11 categories are shown in Fig.~\ref{fig:StruturePrototype}.
In  Table~\ref{tab:HighThou}, we list the chemical formula of materials as well as their space groups contained in each category.
For convenience, we use ``Chemical formula-Magnetic atom'' to name the magnetic atomic structure prototypes. There are 6 categories, including Zr$_{2}$I$_{2}$S$_{2}$-Zr, EuOI-Eu, V$_{3}$N$_{2}$O$_{2}$-V, Cr$_{3}$Te$_{4}$-Cr, Fe$_{3}$GeTe$_{2}$-Fe, and Fe$_{4}$GeTe$_{2}$-Fe that each contains only one material. The remaining 73 materials are divided into five categories.
Note that all the 2DFM materials studied in this work contain only one type of magnetic atom in each structure.
The decoration of nonmagnetic atoms can change the structure dramatically. For example, if the nonmagnetic atoms are taken into account, the 29 original structures of the FeI$_2$-Fe prototype can be further divided into 4 different structures according to their structural similarity.

Among these 79 materials, some have multiple layers of magnetic atoms. Note that the layers we refer to here are those between the layers,
the ions form strong chemical bonds, instead of weak van de Waals bonds.
Among the 11 prototype categories, 4 have a single layer of magnetic atoms, 3 have double magnetic layers, 3 have triple  magnetic layers, and 1 has quadruple layers.
Obviously, the structures of different magnetic layers differ greatly.

The category of FeI$_{2}$-Fe contains 29 2DFM materials, which is the most popular magnetic atomic structure prototype, and the magnetic atoms in this structure prototype reside on the same plane. The original structures have 3 different space groups $P$-$6m2$, $P$-$3m1$, and $P3m1$. The composition types of these structures are AB$_{2}$ and ABC.
The CrI$_{3}$-Cr prototype with only one layer of magnetic atoms contains 14 2DFM materials which possess the $P$-$62m$ and $P$-$31m$ space groups, and adopt AB$_{3}$ and ABC$_{3}$ composition types.  This category includes the famous CrI$_{3}$ compound, which has been experimentally synthesized~\cite{Huang2017Layer},  and investigated intensively.

For VO$_{2}$-V prototype, the magnetic atoms also reside in one plane, and there are $P4/nmm$, $P$-$4m2$, and $P4/mmm$, three different space groups for the original structures. The composition types are AB and AB$_{2}$. The Co$_{2}$F$_{2}$-Co prototype, which has double magnetic atom layers, has 6 2DFM materials and contains AB and ABC two different composition types. Their space groups are all $P$-$3m1$. There are 9 2DFM materials with ABC composition type and $Pmmn$ space group belongs to Cr$_{2}$I$_{2}$S$_{2}$-Cr prototype.

\subsection*{Curie temperatures}

The calculated $T_{\rm C}$s for the 79 2DFM materials are also listed in Table~\ref{tab:HighThou}.
Since $T_{\rm C}$s of some of the 2DFM materials have been measured experimentally, we first compare the calculated $T_{\rm C}$s to the experimental values for these materials.
The calculated $T_{\rm C}$ of monolayer CrI$_{3}$ is 26 K, which is slightly lower than  45 K of the experimental value~\cite{Huang2017Layer} and very close to the other theoretical calculation of 28 K~\cite{Daniele2020High}. The magnetic order at finite temperature can be stabilized when applying a small external magnetic field to Cr$_{2}$Ge$_{2}$Te$_{6}$~\cite{Gong2017Discovery}.
The calculated $T_{\rm C}$ of  Cr$_{2}$Ge$_{2}$Te$_{6}$~\cite{Gong2017Discovery} is 11 K with 0.25 meV/Cr magnetic anisotropic energy is very close to the estimated experimental value, and is slightly lower than the theoretical estimation of 20 K  in Ref.~\cite{Gong2017Discovery}.
It is worth mentioning that the calculated $T_{\rm C}$ of Fe$_{3}$GeTe$_{2}$ is 158 K~\cite{Shen2021magnetic}, which is close to the experimental value of 130 K~\cite{Fei2018Two}. Our calculated $T_{\rm C}$ of monolayer Fe$_{4}$GeTe$_{2}$ is about 207 K which is below 270 K of its bulk phase~\cite{Seo2020Nearly}. We also calculate the $T_{\rm C}$ of monolayer Cr$_{3}$Te$_{4}$, and obtain $T_{\rm C}$=133 K, which is also lower than the experimental $T_{\rm C}$=316 K of the bulk material~\cite{Yamaguchi1972Magnetic}, as expected. Note that in Ref.~\cite{Zhu2018Systematic}, the $T_{\rm C}$ of monolayer Cr$_{3}$Te$_{4}$ was severely overestimated to be 2057 K, much higher than the $T_{\rm C}$ of the bulk  Cr$_{3}$Te$_{4}$.
All these results suggest that our calculated $T_{\rm C}$s of the 2DFM materials are highly reliable.
The exchange interactions calculated by the linear response theory are somehow smaller than those calculated by the energy mapping method \cite{Zhu2018Systematic,Daniele2020High,Kabiraj2020High}, and therefore the calculated $T_{\rm C}$s are also lower.

\begin{figure}
  \centering
  \includegraphics[width=0.45\textwidth]{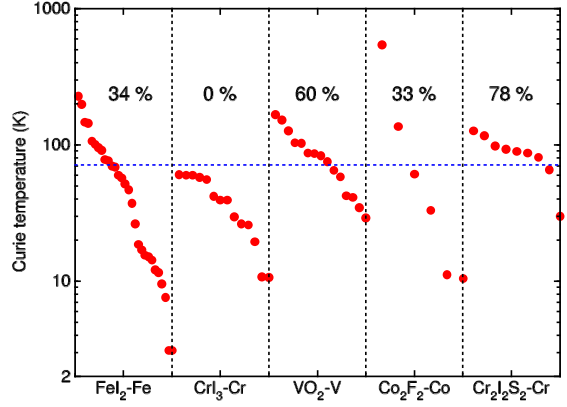}
  \caption{The distribution of $T_{\rm C}$ for 73 2DFM materials in different structure prototypes. The  $T_{\rm C}$s are arranged in descending order.
  Blue dot line marks the average $T_{\rm C}$ for the 73 structures.
  The percentages are the ratios of structures that have $T_{\rm C}$ above the average Curie temperature for each category. }
  \label{fig:TcDistri}
\end{figure}

The distributions of $T_{\rm C}$s of the first five categories of 2DFM materials are shown in Fig.~\ref{fig:TcDistri}. The average $T_{\rm C}$ of the 73 materials is 71 K. The average $T_{\rm C}$ for FeI$_{2}$-Fe, CrI$_{3}$-Cr, VO$_{2}$-V, Co$_{2}$F$_{2}$-Co, and Cr$_{2}$I$_{2}$S$_{2}$-Cr categories are 66 K, 38 K, 84 K, 132 K, and 88 K, respectively. The ratio of structures that have $T_{\rm C}$ higher than the average value (71 K) for the above categories are 34\%, 0\%, 60\%, 33\%, and 78\%, respectively.
The last 6 prototypes in TABLE~\ref{tab:HighThou} have only one 2DFM material in each category. Among them, the $T_{\rm C}$s of
V$_{3}$N$_{2}$O$_{2}$, Cr$_{3}$Te$_{4}$, Fe$_{3}$GeTe$_{2}$, and Fe$_{4}$GeTe$_{2}$ are 71 K, 133 K, 158 K, and 207 K, respectively.

Among the 79 2DFM materials, Co$_{2}$F$_{2}$ has the highest $T_{\rm C}$=541 K.
The  $T_{\rm C}$s of the rest materials are all below room temperature, which is not very surprising.
We find 3 materials whose $T_{\rm C}$ are higher than 200 K, and 4 materials, whose $T_{\rm C}$ are
between 150 K and 200 K. There are 11 materials with $T_{\rm C}$ between 100 K and 150 K.
These results suggest that the $T_{\rm C}$s of 2DFM materials are relative low, compared to the 3D compounds,
and finding room temperature 2DFM materials is quite difficult.
The $T_{\rm C}$s  of previous calculations~\cite{Zhu2018Systematic,Kabiraj2020High} are generally higher than the ones in this work. Especially,
in Ref.\cite{Zhu2018Systematic},  heuristic factors 0.2 - 0.4 are multiplied to obtain reasonable $T_{\rm C}$s.

\begin{table*}[!h]
  \centering
  \caption{The structural prototypes of the 2DFM materials, and the chemical formula, space group of each compound. The
calculated $T_{\rm C}$ (in K) and local magnetic moment (in $\mu_{\rm B}$ per spin) are also shown. 1 L, 2 L, 3 L, and 4L denote that the 2DFM materials have 1, 2, 3, and 4 layers of magnetic atoms, respectively. }
\begin{tabular}{c|cccccccc}
\hline\hline
Prototype&Chemical formula&Space group&$T_{\rm C}$ &$M$ &Chemical formula&Space group& $T_{\rm C}$ &$M$\\
\hline
\multirow{15}{*}{ \shortstack{FeI$_{2}$-Fe\\
(1 L)}}&CrO$_{2}$&$P$-3$m$1&227&2.0&FeCl$_{2}$&$P$-6$m$2&199&4.0\\
&FeBr$_{2}$&$P$-6$m$2&148&4.0&FeTe$_{2}$&$P$-6$m$2&144&1.8\\
&FeO$_{2}$&$P$-3$m$1&107&2.0&CrClI&$P$3$m$1&101&4.0\\
&FeI$_{2}$&$P$-6$m$2&95&4.0&CoO$_{2}$&$P$-3$m$1&91&1.0\\
&VSe$_{2}$&$P$-6$m$2&78&1.0&FeBr$_{2}$&$P$-3$m$1&77&4.0\\
&VS$_{2}$&$P$-6$m$2&70&1.0&VSSe&$P$3$m$1&69&1.0\\
&VTe$_{2}$&$P$-6$m$2&60&1.0&VSeTe&$P$3$m$1&57&1.0\\
&CrBrI&$P$3$m$1&52&4.0&RhI$_{2}$&$P$-3$m$1&47&1.0\\
&VSTe&$P$3$m$1&37&1.1&FeI$_{2}$&$P$-3$m$1&26&4.0\\
&ScCl$_{2}$&$P$-6$m$2&19&1.0&ScBr$_{2}$&$P$-6$m$2&17&1.0\\
&VS$_{2}$&$P$-3$m$1&16&0.5&ScI$_{2}$&$P$-6$m$2&15&1.0\\
&TmI$_{2}$&$P$-3$m$1&14&0.9&VSSe&$P$3$m$1&12&0.7\\
&NbSeTe&$P$3$m$1&12&0.8&NbSTe&$P$3$m$1&10&0.7\\
&NbTe$_{2}$&$P$-6$m$2&8&0.9&YBr$_{2}$&$P$-6$m$2&3&1.0\\
&YCl$_{2}$&$P$-6$m$2&3&1.0&&&&\\
\hline
\multirow{7}{*}{ \shortstack{CrI$_{3}$-Cr\\
(1 L)}}&V$_{2}$Cl$_{6}$&$P$-31$m$&60&2.0&V$_{2}$Br$_{6}$&$P$-31$m$&60&2.0\\
&V$_{2}$I$_{6}$&$P$-31$m$&60&2.0&Ru$_{2}$Cl$_{6}$&$P$-31$m$&58&1.0\\
&Os$_{2}$Cl$_{6}$&$P$-31$m$&56&1.0&Ti$_{2}$Cl$_{6}$&$P$-62$m$&42&1.0\\
&Ti$_{2}$Br$_{6}$&$P$-62$m$&40&1.0&Re$_{2}$Br$_{6}$&$P$-31$m$&40&2.0\\
&Ru$_{2}$Br$_{6}$&$P$-31$m$&30&1.0&Ni$_{2}$I$_{6}$&$P$-31$m$&26&1.0\\
&CrI$_{3}$&$P$-31$m$&26&3.0&Ni$_{2}$Br$_{6}$&$P$-31$m$&20&1.0\\
&Os$_{2}$I$_{6}$&$P$-31$m$&11&1.0&Cr$_{2}$Ge$_{2}$Te$_{6}$&$P$-31$m$&11&3.0\\
\hline
\multirow{8}{*}{ \shortstack{VO$_{2}$-V\\
(1 L)}}&Ni$_{2}$I$_{2}$&$P4/nmm$&166&1.0&VO$_{2}$&$P$-4$m$2&152&1.0\\
&Ni$_{2}$O$_{2}$&$P4/mmm$&127&1.2&Mn$_{2}$Se$_{2}$&$P4/nmm$&104&1.4\\
&Ni$_{2}$Se$_{2}$&$P4/mmm$&103&0.6&Mn$_{2}$S$_{2}$&$P4/nmm$&87&1.3\\
&Ni$_{2}$S$_{2}$&$P4/mmm$&87&0.5&Ni$_{2}$Te$_{2}$&$P4/mmm$&83&0.6\\
&CrI$_{2}$&$P$-4$m$2&75&4.0&Mn$_{2}$Te$_{2}$&$P4/nmm$&65&1.5\\
&Co$_{2}$S$_{2}$&$P4/nmm$&59&0.5&Cr$_{2}$Se$_{2}$&$P4/nmm$&43&0.3\\
&Cr$_{2}$S$_{2}$&$P4/nmm$&41&0.4&Rh$_{2}$S$_{2}$&$P4/nmm$&35&0.3\\
&NiBr$_{2}$&$P$-4$m$2&29&2.0&&&&\\
\hline
\multirow{3}{*}{ \shortstack{Co$_{2}$F$_{2}$-Co\\
(2 L)}}&Co$_{2}$F$_{2}$&$P$-3$m$1&542&2.4&Ni$_{2}$I$_{2}$&$P$-3$m$1&137&1.0\\
&ErHCl&$P$-3$m$1&61&2.6&YbOCl&$P$-3$m$1&33&1.0\\
&Sc$_{2}$Cl$_{2}$&$P$-3$m$1&11&0.8&Sc$_{2}$Br$_{2}$&$P$-3$m$1&11&0.8\\
\hline
\multirow{5}{*}{ \shortstack{Cr$_{2}$I$_{2}$S$_{2}$-Cr\\
(2 L)}}&Mn$_{2}$I$_{2}$O$_{2}$&$Pmmn$&127&4.0&Mn$_{2}$I$_{2}$N$_{2}$&$Pmmn$&117&3.0\\
&V$_{2}$Br$_{2}$O$_{2}$&$Pmmn$&98&2.0&Mn$_{2}$Cl$_{2}$N$_{2}$&$Pmmn$&93&3.0\\
&Cr$_{2}$I$_{2}$Se$_{2}$&$Pmmn$&90&3.0&Cr$_{2}$I$_{2}$S$_{2}$&$Pmmn$&88&3.0\\
&Cr$_{2}$Br$_{2}$S$_{2}$&$Pmmn$&81&3.0&Cr$_{2}$Cl$_{2}$S$_{2}$&$Pmmn$&66&3.0\\
&HoSI&$Pmmn$&30&4.0&&&&\\
\hline
\multirow{2}{*}{\shortstack{Zr$_{2}$I$_{2}$S$_{2}$-Zr\\
(1 L)}}&\multirow{2}{*}{Zr$_{2}$I$_{2}$S$_{2}$}&\multirow{2}{*}{$Pc$}&\multirow{2}{*}{5}&\multirow{2}{*}{0.8}&&&&\\
&&&&&&&&\\
\hline
\multirow{2}{*}{\shortstack{EuOI-Eu\\
(2 L)}}&\multirow{2}{*}{EuOI}&\multirow{2}{*}{$P4/nmm$}&\multirow{2}{*}{35}&\multirow{2}{*}{7.4}&&&&\\
&&&&&&&&\\
\hline
\multirow{2}{*}{\shortstack{V$_{3}$N$_{2}$O$_{2}$-V\\
(3 L)}}&\multirow{2}{*}{V$_{3}$N$_{2}$O$_{2}$}&\multirow{2}{*}{$P$-6$m$2}&\multirow{2}{*}{71}&\multirow{2}{*}{1.0}&&&&\\
&&&&&&&&\\
\hline
\multirow{2}{*}{\shortstack{Cr$_{3}$Te$_{4}$-Cr\\
3 L}}&\multirow{2}{*}{Cr$_{3}$Te$_{4}$}&\multirow{2}{*}{$P$-3$m$1}&\multirow{2}{*}{133}&\multirow{2}{*}{3.3}&&&&\\
&&&&&&&&\\
\hline
\multirow{2}{*}{\shortstack{Fe$_{3}$GeTe$_{2}$-Fe\\
(3 L)}}&\multirow{2}{*}{Fe$_{3}$GeTe$_{2}$}&\multirow{2}{*}{$P$-6$m$2}&\multirow{2}{*}{158}&\multirow{2}{*}{1.5}&&&&\\
&&&&&&&&\\
\hline
\multirow{2}{*}{\shortstack{Fe$_{4}$GeTe$_{2}$-Fe\\
(4 L)}}&\multirow{2}{*}{Fe$_{4}$GeTe$_{2}$}&\multirow{2}{*}{$P$-3$m$1}&\multirow{2}{*}{207}&\multirow{2}{*}{2.1}&&&&\\
&&&&&&&&\\
\hline
\hline
\end{tabular}
  \label{tab:HighThou}
\end{table*}

\section*{Discussion}

\subsection*{Structure characteristics}

\begin{figure}
  \centering
  \includegraphics[width=0.45\textwidth]{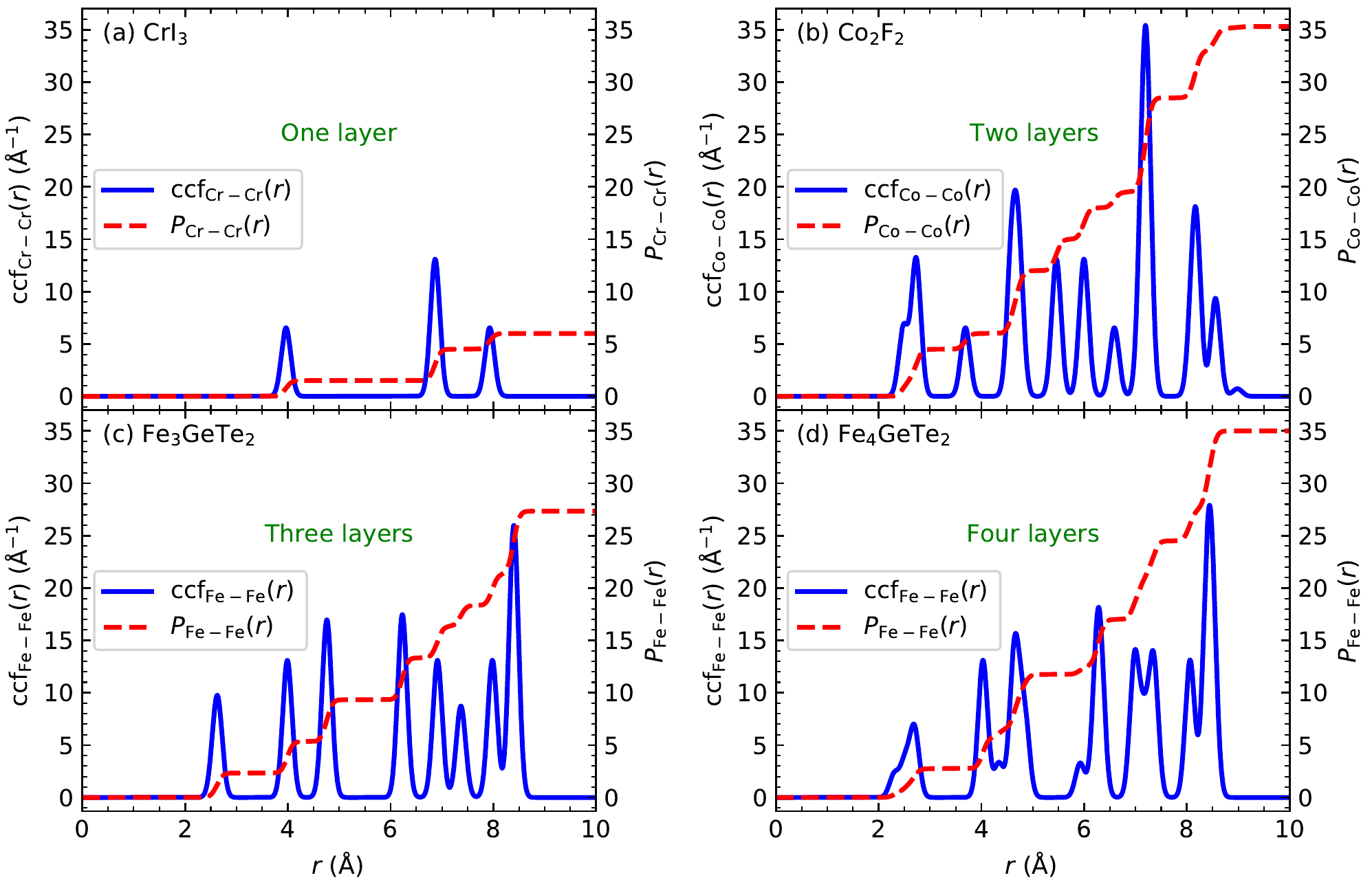}
  \caption{The $ccf_{ij}(r)$ and their integral $P_{ij}(r)$ for (a) CrI$_{3}$, (b) Co$_{2}$F$_{2}$, (c) Fe$_{3}$GeTe$_{2}$, and (d) Fe$_{4}$GeTe$_{2}$. }
  \label{fig:CcfP}
\end{figure}

It is important to understand the relation between the structure and $T_{\rm C}$.
We have checked that the cation-anion-cation angles of all the 79 2DFM materials are close to 90$\degree$, which
obey the Goodenough-Kanamori rule~\cite{Goodenough1958An,Junjiro1959Superexchange} for the FM exchange interactions.
For example, the bond angles of Cr-O-Cr for CrO$_{2}$ and Fe-Cl-Fe angle for FeCl$_{2}$ are 97.53$^\circ$  and 85.49$^\circ$ respectively.

Naively, one could expect that having more neighboring magnetic atoms may lead to higher $T_{\rm C}$ for a material.
The coordination characterization function (CCF) ~\cite{Su2017Construction} defined in Sec. \ref{sec:prototype} can reflect the coordination character of the structure, and the abscissa of the peak of CCF corresponds to the length of a atomic pair. The integral of CCF,
\begin{equation}
P_{ij}(r)= \int_{0}^{r}{\rm ccf}_{ij}(r')\,dr',
	\label{func:itccf}
\end{equation}
 characterizes the average number of (magnetic) atomic pairs
within a certain range $r$,  where ${\rm ccf}_{ij}(r')$ is the CCF for the $i$-th and $j$-th type of element of the structure.

The ${\rm ccf}_{ij}(r)$ (red solid lines) and $P_{ij}(r)$ (blue dashed lines) of four representative structures, CrI$_3$, Co$_{2}$F$_{2}$, Fe$_{3}$GeTe$_{2}$, and Fe$_{4}$GeTe$_{2}$, from different structure prototypes are shown in Fig.~\ref{fig:CcfP}. These structures have different numbers of magnetic atomic layers. CrI$_3$ has quite low $T_{\rm C}$, which is about 26 K, whereas other components have much higher $T_{\rm C}$. Especially,  the $T_{\rm C}$ of Co$_{2}$F$_{2}$ is about 543 K, which is the highest among the 79 2DFM materials. Fe$_{3}$GeTe$_{2}$, and Fe$_{4}$GeTe$_{2}$ also have relative high $T_{\rm C}$. By comparing the ${\rm ccf}_{ij}(r)$ and $P_{ij}(r)$ of different structures, obviously CrI$_3$ has much less neighboring magnetic atoms within given the range than other three components, due to two reasons.
The first factor that affect ${\rm ccf}_{ij}(r)$  and $P_{ij}(r)$ is the length between the nearest magnetic atoms, $d_{\rm min}$.
The smaller $d_{\rm min}$ will increase the neighboring magnetic atoms, and it may also lead to larger exchange interactions between magnetic atoms, which therefore leads to higher $T_{\rm C}$.
The  $d_{\rm min}$  of Co$_{2}$F$_{2}$  is very small ($\sim$ 2.48 \AA), due to the small radii of F atoms.
The $d_{\rm min}$s  of Fe$_{3}$GeTe$_{2}$ and Fe$_{4}$GeTe$_{2}$ are also
quite small $\sim$ 2.5 \AA.  In contrast, the $d_{\rm min}$ of CrI$_3$ is about 3.97 \AA, which is much larger than those of the other three compounds.
The ${\rm ccf}_{ij}(r)$  and $P_{ij}(r)$  are also related to the number of magnetic atomic layers for the materials. For example, CrI$_3$ has only a single layer of magnetic atoms,
whereas Co$_{2}$F$_{2}$ has two magnetic layers and Fe$_{3}$GeTe$_{2}$, Fe$_{4}$GeTe$_{2}$ have 3 and 4 magnetic layers respectively,
which therefore have more neighboring magnetic atoms.
From Table I, one may find that the 2DFM materials with multiple magnetic atomic layers are more likely to have higher $T_{\rm C}$.
In fact, the $T_{\rm C}$s of the most compounds of a single magnetic layer are relatively low, except CrO$_2$ and FeCl$_2$.

The magnetization intensity is often used to evaluate bulk magnetic materials. For 2DFM materials, we define the areal magnetic moment density to measure their macroscopic magnetization,
\begin{equation}
	\mathbf{M}_{\rm 2D} = \frac{\mathbf{m}_{\rm cell}}{s_{\rm cell}},
	\label{func:mag_density}
\end{equation}
where  $\mathbf{m}_{\rm cell}$
 is the total magnetic moment of the unit cell, and $s_{\rm cell}$ is the area of the 2D unit cell.
2DFM materials with larger $\mathbf{M}_{\rm 2D}$ have stronger magnetism and are more favorable for applications.
Obviously, the 2DFM material with denser magnetic atomic mesh, multiple magnetic layers, and larger local magnetic moments would have larger $\mathbf{M}_{\rm 2D}$. For example, the $\mathbf{M}_{\rm 2D}$ for Co$_2$F$_2$, Fe$_4$GeTe$_2$, and Fe$_{3}$GeTe$_{2}$ are 0.74, 0.59, and 0.33 $\mu_{\rm B}$/\AA $^{2}$, respectively. They are much larger than that of CrI$_{3}$ 0.15 $\mu_{\rm B}$/\AA $^{2}$. This is because Co$_2$F$_2$ has two layers of magnetic atoms, which are densely packed. Fe$_4$GeTe$_2$ and Fe$_{3}$GeTe$_{2}$ have 4 and 3 layers of Fe atoms.
In contrast, CrI$_{3}$ has only one layer of magnetic atoms, with relatively larger Cr-Cr pair lengths. Interestingly, Co$_2$F$_2$ possesses both the highest $T_{\rm C}$ and the highest $\mathbf{M}_{\rm 2D}$ among the 79 2DFM materials, showing that it is a promising material.

\begin{figure}[tb!]
  \centering
  \includegraphics[width=0.45\textwidth]{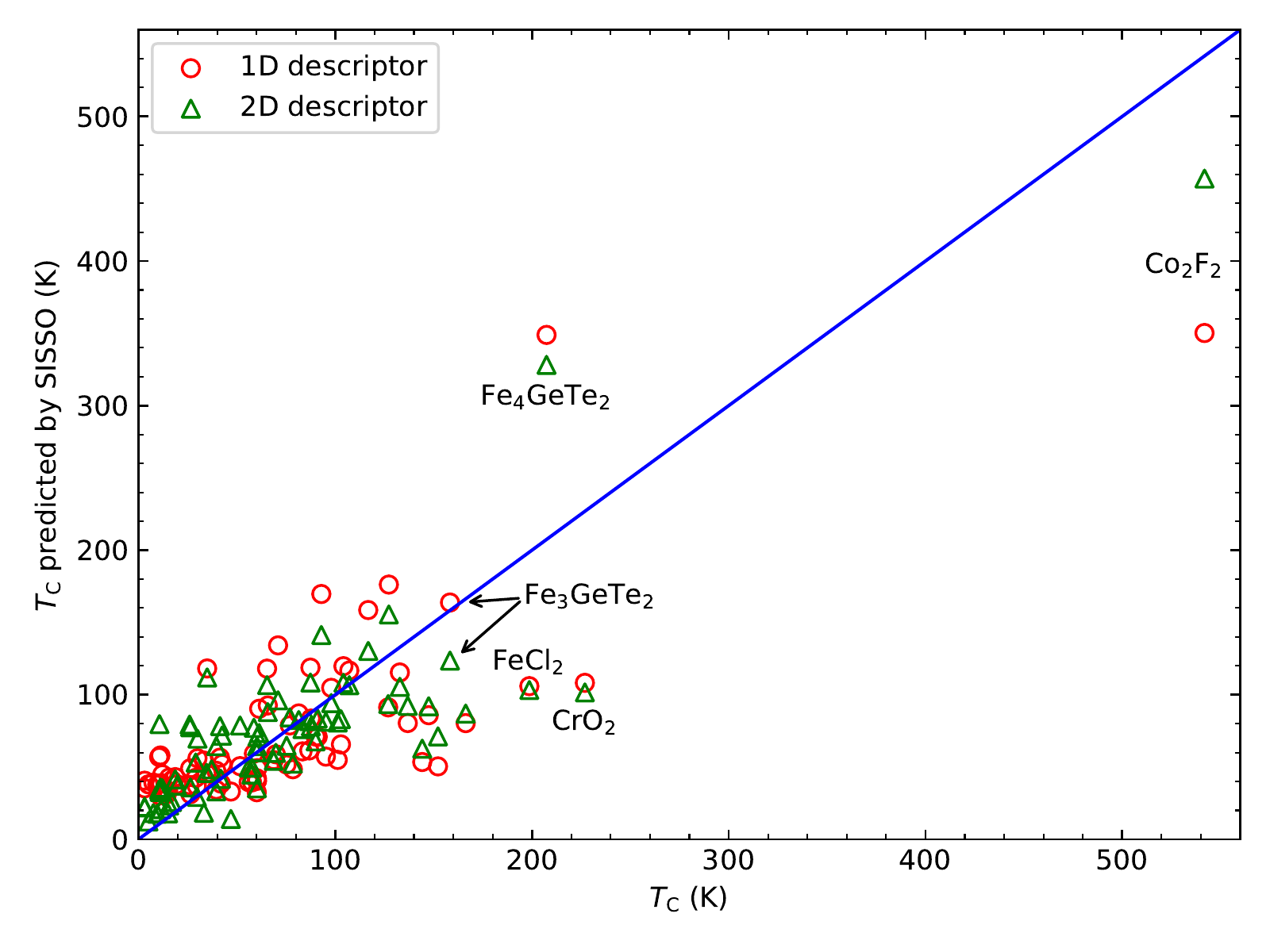}
  \caption{The $T_{\rm C}$ predicted by SISSO vs. the  $T_{\rm C}$ calculated via the high-throughput first-principles calculations. The red circles are the results of 1D descriptor, whereas the
  green triangles are the results of  2D descriptor. }
  \label{fig:SissoR7}
\end{figure}

\subsection*{SISSO analysis of Curie temperature}

To dig out the factors that determine the $T_{\rm C}$ of the 2DFM materials, we resort to the recently developed SISSO method
which may capture the underlying mechanisms of materials' properties using small data sets~\cite{fan2008sure,ouyang2018sisso}.
We use three parameters to construct the model, including the magnetic moment ($M$ in $\mu_{\rm B}$ per spin), the minimum magnetic atomic distance ($d_{\rm min}$, in \AA),
and the integral of CCF ($P$) as typical features to predict the $T_{\rm C}$ of the above 2DFM materials. Among the three parameters, $M$ is an important physical feature of the magnetic atoms, whereas $d_{\rm min}$ and $P$ are used to characterize the structures of the magnetic atoms. The cutoff radius of $P_{ij}(r)$ is set to be 7.0 \AA \  to calculate the $P$ of the above 79 magnetic atomic structures.

Figure~\ref{fig:SissoR7} illustrates the $T_{\rm C}$ predicted by SISSO vs. the {\it ab initio} calculated $T_{\rm C}$.
The one dimensional (1D) descriptor~\cite{ouyang2018sisso} fitting gives,
\begin{equation}
T_{\rm C} = 27.88 + 110.03 \times \frac{M \times P}{d_{\rm min}^{3}} \, ,
\end{equation}
and 2D descriptor gives,
\begin{equation}
T_{\rm C} = 139.33 + 0.0078 \times \frac{M^{3} \times P^{4}}{d_{\rm min}^{4}} - 1.44 \times \frac{P^{\frac{1}{4}} \times d_{\rm min}^{3}}{M^{\frac{1}{2}}}.
\end{equation}
The root mean square error (RMSE) for 2D descriptor is 38 K, which is somehow smaller than that of the 1D descriptor of 46 K.
According to the results of SISSO,  it is clear that $T_{\rm C}$ has a positive correlation with $M$ and $P$, and negative correlation with $d_{\rm min}$,
which is consistent with the analysis in the previous paragraphs.

The SISSO method correctly produces that  Co$_{2}$F$_{2}$ has very high $T_{\rm C}$,
because Co$_{2}$F$_{2}$ meets all the above beneficial conditions for high $T_{\rm C}$.
The $d_{\rm min}$ for Co$_{2}$F$_{2}$ is only 2.48 \AA , and the next nearest neighbor is as short as 2.73 \AA.
The exchange interaction between the nearest neighbor Co-Co pairs in Co$_{2}$F$_{2}$ is calculated to be -35.43 meV, and the exchange interaction between
the next nearest neighbor Co-Co pair is -15.28 meV.
Furthermore, there are about 20 Co-Co pairs with a mutual distance smaller than 7.0 {\AA} for each Co atom, as shown in Fig.~\ref{fig:CcfP}(d).
All these physical quantities suggest that it should have a high $T_{\rm C}$.

On the other hand, the $T_{\rm C}$s of the compounds in the CrI$_{3}$-Cr structural prototype are quite low. For example, CrI$_{3}$ has a low $T_{\rm C}$ 26 K. It has only one magnetic atomic layer and for each Cr atom, there are only 5 Cr-Cr pairs within 7.0 \AA. The $d_{\rm min}$ is as large as 3.97 \AA.
The situations for other structures of CrI$_{3}$-Cr structural prototype are similar.

However, the magnetic interactions have very complicated mechanisms, and the $T_{\rm C}$s are determined by many factors. One would not expect that the simple SISSO models can predict very accurate $T_{\rm C}$s for the 2DFM materials. For example, while the SISSO method predicts the $T_{\rm C}$s  of Fe$_{3}$GeTe$_{2}$  rather well, it somehow over estimates the  $T_{\rm C}$  of Fe$_{4}$GeTe$_{2}$. They also under estimate the $T_{\rm C}$ of FeCl$_2$ and CrO$_2$. Nevertheless, it gives correct trends of $T_{\rm C}$ respect to the structure of the 2DFM materials, which may provide useful guidance to further searching of high  $T_{\rm C}$ 2DFM materials.

Finally, we note that the PBE functional used in this work is suitable for large-scale, fast screening of the materials,
but it is known to have some deficiencies. Once we find the promising candidate materials, we can use
advanced techniques\cite{lee2020role,menichetti2019electronic,ke2021electron,wu2019physical} to obtain more accurate results for the electronic and magnetic structures.
Furthermore, the 2D structures studied in this work are taken from existing databases. It is possible to investigate
a wider range of 2D magnetic materials via crystal structure prediction methods. \cite{wang2012effective,sharan2022intrinsic}

In this work, we carry out high-throughput first-principles computations to search for potential high $T_{\rm C}$ 2DFM materials. We identify 79 2D materials that have robust FM ground state and calculate their $T_{\rm C}$s. Among the 79 2DFM materials,  Co$_{2}$F$_{2}$ has the highest $T_{\rm C}$=541K, well above room temperature. There are three 2D materials that also have  relative high $T_{\rm C}$, including CrO$_{2}$ (226 K), FeCl$_{2}$ (198 K), and Fe$_{4}$GeTe$_{2}$ (207 K).
We perform SISSO method to analyze the relations between the $T_{\rm C}$ and the structure of the 2DFM materials.
The results suggest that the 2DFM materials with smaller distance between the magnetic atoms $d_{\rm min}$, larger local magnetic moments
and more neighboring magnetic atoms are more likely to have higher $T_{\rm C}$.
Our research is instructive to the discovery of new 2DFM materials with high $T_{\rm C}$.

\section*{Methods}
\label{sec:method}

\subsection*{Workflow}

\begin{figure}
  \centering
  \includegraphics[width=0.48\textwidth]{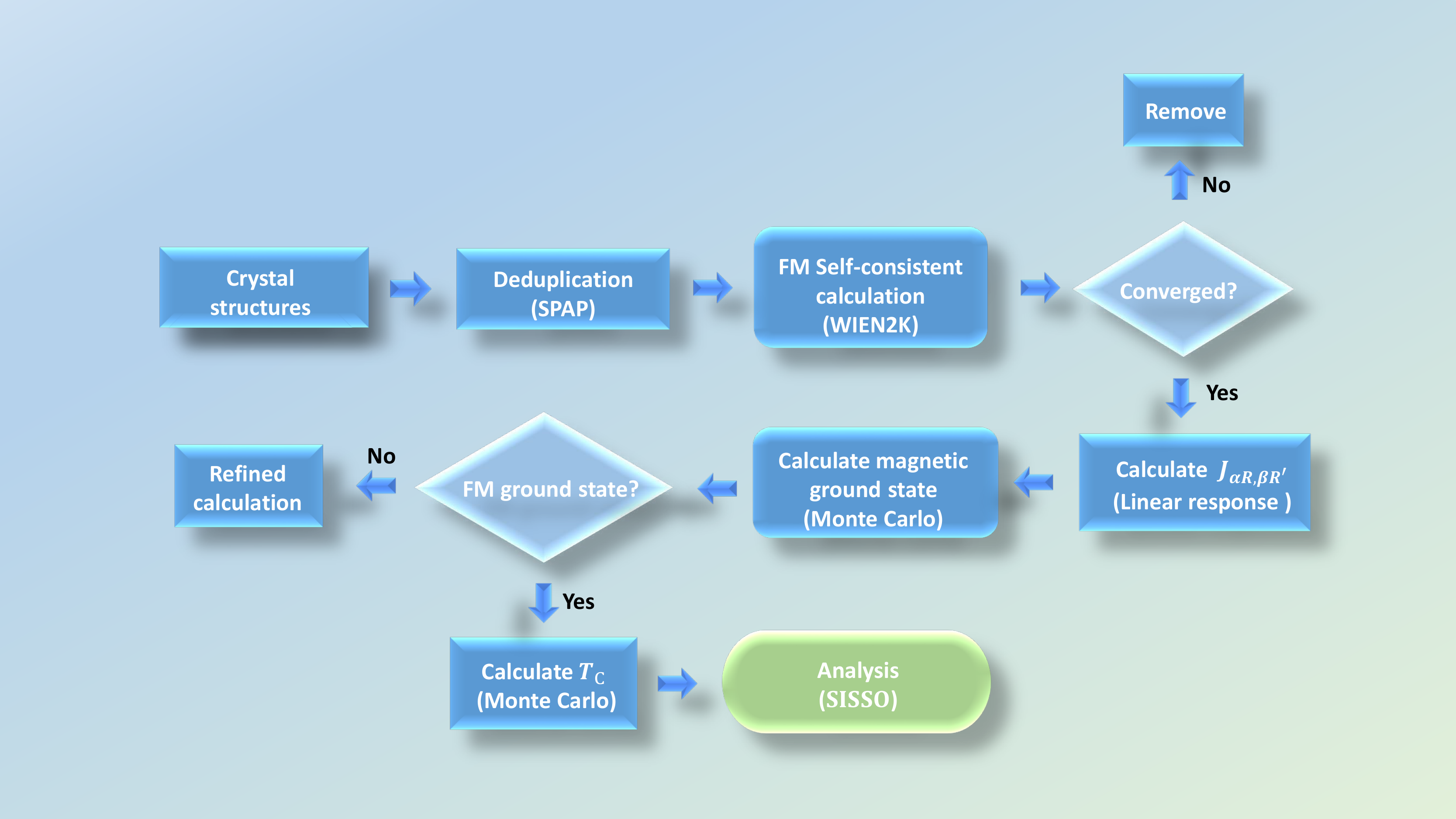}
  \caption{The workflow of the high-throughput computation of the 2D magnetic materials. }
  \label{fig:WorkFlow}
\end{figure}

The work flow of our high-throughput computations is illustrated in Fig.~\ref{fig:WorkFlow}. This high-throughput computing process is highly automated and almost no empirical parameters are introduced. Once the basic 2D magnetic crystals are collected, we run the  Structure Prototype Analysis Package (SPAP)
code~\cite{Su2017Construction} to cluster structures according to structural similarity and label the structures with corresponding structure prototypes. We then perform self-consistent first-principle calculations,
and some non-convergent structures will be removed at this step. The magnetic exchange interactions $J_{\alpha \mathbf{R}, \beta \mathbf{R}^{'}}$ are calculated  by a linear response method~\cite{Wan2006Calculation,Wan2009Calculated}. Based on the calculated magnetic exchange parameters, the magnetic ground state of the system is simulated by the Monte Carlo method at $T \approx$ 0 K. If the ground state is ferromagnetic,  the  $T_{\rm C}$ will be calculated. We carry out further analysis based on these results.

\subsection*{Structure prototype analysis}
\label{sec:prototype}

The properties of the materials are determined by their constituting elements and the residing structure. A group of materials having similar structures tend to possess similar properties. In order to find out the relationship between the $T_{\rm C}$ (and other magnetic properties) of the 2DFM materials and their structures, we categorize the 2DFM materials according to their structural similarity. We use the SPAP code to classify these structures into corresponding structure prototypes, and details of the methods are given in
Ref.~\cite{Su2017Construction}. Different from the similarity comparisons of bulk crystal structures, which needs to normalize two structures to the same volumetric atom number density, for 2D materials, we need to construct an areal atom number density normalization coefficient,
\begin{equation}
	c_{\rm 2D} =\sqrt{\frac{S_{A}N_{B}}{S_{B}N_{A}}},
	\label{func:c_2D}
\end{equation}
where $S_A$, $S_B$, $N_A$, and $N_B$ denote the area of the 2D lattice of structure $A$ and $B$, and the number of atoms in the 2D cell of structure $A$ and $B$, respectively. We multiply the 3 lattice vectors of structure $B$ by $c_{\rm 2D}$. Note that the fractional coordinates of the atoms of the structure $B$ should be kept unchanged. The manipulated structure $B$ is adjusted to the same areal atom number density with that of structure $A$. The normalization helps us to identity similar structures with different scales. We use the CCF~\cite{Su2017Construction} to measure the distance between normalized structures,
which is defined as follows, for pairs of atomic types $i$ and $j$,
\begin{equation}
	{\rm ccf}_{ij}(r)=
(1-{1 \over 2}\delta_{ij})\frac{1}{N} \sum\limits_{n_{i}} \sum\limits_{n_{j}}f(r_{n_{i}n_{j}})\sqrt{\frac{a_{\rm pw}}{\pi}}e^{-a_{\rm pw}(r-r_{n_{i}n_{j}})^2},
		\label{func:ccf}
\end{equation}
where $N$ is the total number of atoms in the cell. $n_i$ runs over all atoms of the $i$-th type within the unit cell, whereas, $n_j$ runs over all atoms of the $j$-th type within
the extended cell. $f(r_{n_{i}n_{j}})$ is the weighting function for different interatomic distances, whereas $r_{n_{i}n_{j}}$ is the interatomic distance within the cutoff radius. $a_{\rm pw}$ (usually 60.0 \AA$^{-2}$) is a parameter that controls the width of the normalized Gaussian smearing function.

We assign similar structures with the same prototype name with the help of three norms: (i) the same composition type and total number of atoms in the conventional cell; (ii) equivalent three-dimensional crystal symmetry; and (iii) the distance between two structures being below a threshold. In this work, we take 0.075 as the threshold.
After the structural classification by SPAP, the structures are rechecked visually to guarantee that similar structures are labeled with the same prototype name.

\subsection*{Density functional calculation}

The structural, electronic, and magnetic properties of the 2D magnetic materials are studied via density functional theory (DFT) by the WIEN2K code, which is based on a full-potential  linearized augmented plane wave method (FP-LAPW)~\cite{Blaha2001WIEN2k}. We adopt the Perdew-Burke-Ernzerhof (PBE) form generalized gradient approximation (GGA) of the exchange-correlation functional~\cite{Perdew1996Generalized}. For all calculated materials, 1000 Brillouin $k$-points are used, and the number of $k$-points are automatically assigned according to the shape of the unit cell. The plane-wave cutoff $RK_{max}$ = 7 is used.

\subsection*{Magnetic exchange interactions}

In previous works~\cite{Zhu2018Systematic,Torelli2019High,Daniele2020High,Kabiraj2020High}, the magnetic exchange interactions are usually calculated via an energy mapping method, i.e., the total energies of different spin configurations are calculated, and the exchange interactions are fitted to the energies of different spin configurations. There are several disadvantages of this method. First, this method is not user-friendly for the high-throughput calculations, since the spin configurations have to be chosen by hand for each material, with caution. Different choices of the spin configurations may lead to large discrepancies in the results.
This is because, for some spin configurations, the electrons are forced to occupy higher energy orbitals, and in these cases,
the total energies can not be fitted very well by the Heisenberg model, because an additional electron band (the so-called Stoner) energies have to be considered.
The exchange interactions calculated by the energy mapping method may thus be overestimated and
this might be one of the reasons that some of the previous works significantly overestimated the $T_{\rm C}$\cite{Zhu2018Systematic}.
Furthermore, for some materials, the long-range exchange interactions are important for $T_{\rm C}$~\cite{Kabiraj2020High}. To take account of the long-range magnetic interactions, very large supercells are required to obtain the exchange interactions.

In this work, we calculate the magnetic interactions via a first-principle linear-response approach~\cite{Wan2006Calculation,Wan2009Calculated},
where the magnetic exchange interaction is calculated as:
 \begin{equation}
J_{i\mathbf{R}, j\mathbf{R}^{\prime}}  =  \sum_{\mathbf{q}} \sum_{\mathbf{k} n n^{\prime}} \frac{f_{n\mathbf{k}}-f_{n^{\prime}\mathbf{k}+\mathbf{q} }}{\epsilon_{n\mathbf{k}}-\epsilon_{n^{\prime}\mathbf{k}+\mathbf{q}}} \langle \psi_{n\mathbf{k} }\left|\left[\sigma \times \mathbf{B}_{i}\right]\right| \psi_{n^{\prime}\mathbf{k}+\mathbf{q}} \rangle \times  \langle\psi_{ n^{\prime}\mathbf{k}+\mathbf{q}}\left|\left[\sigma \times \mathbf{B}_{j}\right]\right| \psi_{n\mathbf{k} } \rangle e^{i \mathbf{q} \cdot \left(\mathbf{R}-\mathbf{R}^{\prime}\right)} \, ,
\end{equation}
where, $\epsilon_{n\mathbf{k}}$ is the one-electron band energy, $\sigma$ are the Pauli matrices.
$\psi_{n\mathbf{k}}$ is the Kohn-Sham wave function, $\mathbf{B}_{i}$ is
the local magnetic field, $\mathbf{R}$ is the unit cell index, whereas $i$ is the atom in the $\mathbf{R}$-th unit cell.
This technique has been successfully applied to a wide variety of complex magnetic materials~\cite{Wan2009Calculated, Wan2011Mechanism,Wang2019Calculated,Bo2020Calculated,Shen2021magnetic}.
Details about the method can be found in Ref.~\cite{Wan2006Calculation,Wan2009Calculated}.
In this work, we calculate all the eligible exchange interactions $J_{i \mathbf{R}, j \mathbf{R^{'}}}$, i.e., the exchange interactions between the $i$-th spin in the $\mathbf{R}$-th unit cell and the $j$-th spin in the $\mathbf{R}^{'}$-th unit cell,
 within the range of $|\mathbf{R_{\alpha}^{'}} - \mathbf{R}_{\alpha}|  \leq 3$, where $\alpha$ = $a$, $b$, $c$.

\subsection*{Replica-exchange Monte Carlo simulation}

The magnetic phase diagrams and the $T_{\rm C}$s are simulated via the following anisotropic Heisenberg-like Hamiltonian,
\begin{equation}
	H = \sum_{i\mathbf{R},j\mathbf{R}^{'}}J_{i \mathbf{R},j\mathbf{R}^{'}}{\bf S}_{i\mathbf{R} } \cdot {\bf S}_{j \mathbf{R}^{'}} + \sum_{i} A({\bf S}_{i}^{z})^{2},
	\label{func:Hamiltonian}
\end{equation}
where ${\bf S}_{i\mathbf{R}}$ is the $i$-th spin in the $\mathbf{R}$-th unit cell.
 The magnetic anisotropy energy $A$ is important to stabilize the ferromagnetism in 2D.
Because the magnetic anisotropy energy $A$ is very sensitive to the calculation parameters, it is still quite difficult to obtain highly accurate and reliable values in the high-throughput DFT calculations. Fortunately, we find that $T_{\rm C}$ is not very sensitive to the value of $A$, as demonstrated in our
previous works~\cite{Shen2021magnetic}. For example, the $T_{\rm C}$ of CrI$_{3}$ with the magnetic anisotropic energy $A$=0.25 meV for each Cr ion is 26 K, whereas the $T_{\rm C}$ only increases to 30 K when $A$=1.00 meV is used.
Therefore we set a reasonable value $A$=0.25 meV per spin for all materials, which is enough for our purpose.
We simulate the above spin Hamiltonian via the replica-exchange Monte Carlo method~\cite{Swendsen1986Replica,Bruno2001Absence,Cao2009First},
in which one simulates $M$ replicas each at a different temperature $T$ covering a range of interest, allowing configurational exchange between the replicas.
The simulations are performed on a 36$\times$36 lattice with periodic boundary conditions. We first perform the simulations at temperatures ranging from 5 to 640 K using 120 replicas. If no FM phase transition is found, we then perform simulation at temperatures ranging from 0.2 to 20 K. We discard the first 2$\times$10$^5$ sweeps,
when computing the equilibrium properties. Sample averages are accumulated over 2$\times$10$^5$ sweeps.

\subsection*{SISSO analyses}

SISSO is an approach for discovering descriptors for materials’ properties, which can tackle huge and possibly strongly correlated feature spaces, and converges to the optimal solution from a combination of features relevant to the materials’ target property. The outcome of SISSO is explicit, analytic functions consisted of basic physical quantities. A detailed description of the method can be found in
Refs.~\cite{fan2008sure,ouyang2018sisso}. In this work, the SISSO code from github.com/rouyang2017/SISSO is used to perform the analyses.

\section*{DATA AVAILABILITY}

All data generated and/or analyzed during this study
are included in this article.

\section*{CODE AVAILABILITY}

The SPAP code is available at https://github.com/chuanxun/StructurePrototypeAnalysisPackage. The homemade Monte Carlo simulation code
is available upon request to He, Lixin (helx@ustc.edu.cn). The linear-response code to calculate the exchange interactions
is available upon request to Wan, Xiangang (xgwan@nju.edu.cn).

\section*{Acknowledgements}
We appreciate Prof. Wan, Xiangang for generously allowing us to use the linear-response code.
This work was funded by the Chinese National Science
Foundation Grant Number 12134012. The numerical calculations were done on the USTC HPC facilities.

\section*{AUTHOR CONTRIBUTIONS}

Z. Shen and C. Su performed the calculations.
All authors analyzed the results and wrote the manuscript. L. He  conducted the project.

\section*{Competing interests:}
The authors declare no competing interests.


\end{spacing}
\end{document}